# Spectroscopic, Structural, and Theoretical Studies of Halide Complexes with a Urea-based Tripodal Receptor

*Avijit Pramanik[1], Douglas R. Powell[2], Bryan M. Wong[3], and Md. Alamgir Hossain*[1]*

[1]Department of Chemistry and Biochemistry, Jackson State University, Jackson, MS 39212,

[2]Department of Chemistry and Biochemistry, University of Oklahoma, Norman, OK 73019

[3]Materials Chemistry Department, Sandia National Laboratories, Livermore, CA 94551

Email: alamgir.hossain@jsums.edu

ABSTRACT: A urea-based tripodal receptor **L** substituted with *p*-cyanophenyl groups has been studied for halide anions by $^1$H NMR spectroscopy, density functional theory (DFT) calculations and X-ray crystallography. The $^1$H NMR titration studies suggest that the receptor forms a 1:1 complex with an anion, showing the binding trend in the order of fluoride > chloride > bromide > iodide. The interaction of fluoride anion with the receptor was further confirmed by 2D NOESY and $^{19}$F NMR spectroscopy in DMSO-$d_6$. DFT calculations indicate that the internal halide anion is held by six NH···X interactions with **L**, showing the highest binding energy for the fluoride complex. Structural characterization of the chloride, bromide, and silicon hexafluoride complexes of [**LH**$^+$] reveals that the anion is externally located via hydrogen bonding interactions. For the bromide or chloride complex, two anions are bridged with two receptors to form a centrosymmetric dimer, while for the silicon hexafluoride complex, the anion is located within a cage formed by six ligands and two water molecules.

**Introduction**

Anions are ubiquitous in nature and play a key role in chemistry and biology.[1] Therefore, anion recognition with synthetic receptors remains an active area of research in supramolecular chemistry.[2,3] Early research in this area has focused primarily on polyamine-based receptors that require protonation to bind an anion.[4-7] In order to overcome this limitation, researchers have started using neutral molecules functionalized with amide,[8-11] thioamide,[12,13] urea,[14-16] thiourea,[17,18] pyrrole,[19-21] and indole[22-25] groups that can readily form H-bonds with an anion regardless of solution pH. In particular, the electron withdrawing nature of the oxygen atom in the urea-based molecule can result in the formation of two hydrogen bonds with an anionic guest, providing directional binding modes (Scheme 1). For example, a simple dimethyl urea receptor containing a single urea group reported by Hamilton *et al.* was shown to bind an acetate ($K$ = 45 M$^{-1}$) in DMSO.[26] Attaching the urea group to two 4-nitrophenyl groups, Fabbrizzi *et al.* synthesized a bis(4-nitrophenyl) urea receptor which was shown to form a 1:1 complex with a variety of anions, showing a high affinity for fluoride ($K$ = 2.40 x 10$^7$ M$^{-1}$) in CH$_3$CN.[27] Albrecht *et al.* a reported a quinoline-based receptor containing both amide and urea groups that was found to complex halides in CHCl$_3$, showing a high affinity for fluoride ($K$ = 1.44 x 10$^5$ M$^{-1}$).[28] Gale *et al.* obtained a urea-based receptor with attached indole groups and isolated a crystal with carbonate in which the anion species was surrounded by two receptors with both indole and urea NH functional groups.[29] Johnson *et al.* characterized a dipodal urea based on rigid acetylene groups with a central pyridine framework, which after protonation binds a chloride in a pentadentate fashion, forming a five-coordinate chloride complex.[30] Martinez-Máñez *et al.* prepared colorimetric sensors by attaching 4-nitroazobenzene to an acyclic urea appended to a dye, showing a complex with atmospheric CO$_2$ in the presence of fluoride ion. The resulting carbonate complex was formed due to the deprotonation of the NH groups by fluoride ion in water.[31]

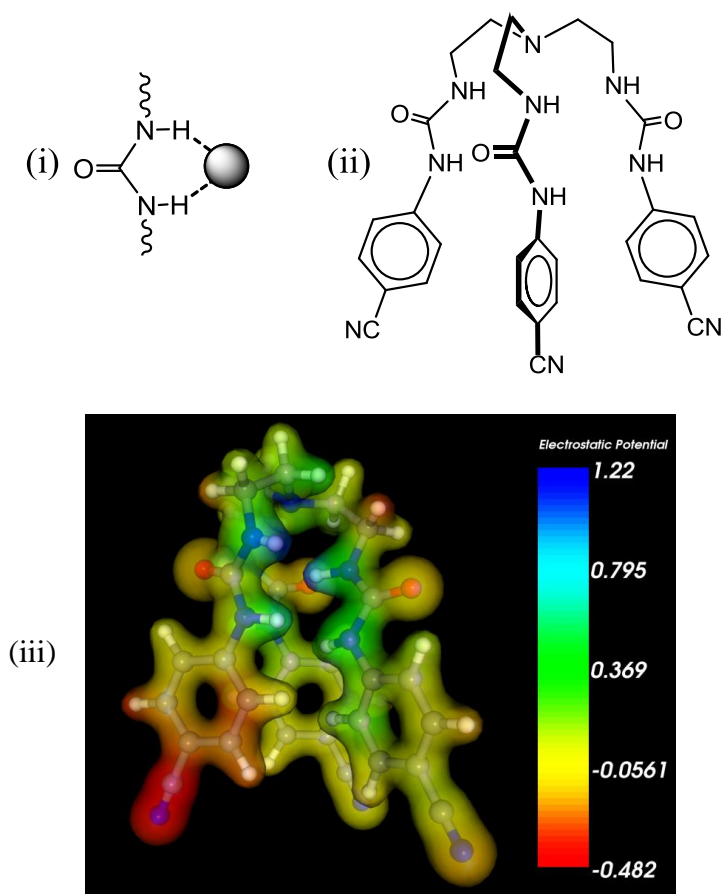

**Scheme 1.** Schematic representation of (i) a urea unit and its directional binding mode for an anion, (ii) urea based tripodal receptor (**L**), and (iii) electrostatic potential map for **L** calculated at the M06-2X/6-31G(d,p) level of theory (red = negative potential, blue = positive potential).

A number of synthetic receptors based on tris(2-aminoethyl)amine (tren) have been reported in the literature, which are primarily limited to polyamines.[32-37] Recently, several groups have taken advantage of using this framework for synthesizing urea-based neutral hosts as effective receptors to complex an anion with multiple H-bonds.[38-44] For example, Custelcean *et al.* reported a tripodal urea substituted with *m*-cyanophenyl groups that formed a silver-based MOF in the presence of $Ag_2SO_4$, where a doubly charged sulfate was encapsulated with twelve hydrogen bonds.[38] Wu *et al.* reported a multiply-coordinated sulfate complex with a tripodal urea substituted with 3-pyridyl groups.[39] Ghosh *et al.* reported a pentafluorophenyl-substituted tripodal urea encapsulating a fluoride anion with six NH bonds.[40] Very recently, Gale and coworkers reported a series of fluorinated tren-based ureas and thioureas which have been shown to function as anticancer agents through transmembrane transport mechanism of anions in vitro.[44]

We have previously reported a tripodal receptor, **L**, with three *p*-cyanophenyl groups as electron withdrawing substituents, showing high selectivity for sulfate and hydrogen sulfate over other oxoanions.[45] It was hoped that the introduction of this group would enhance the acidity of the attached NH groups, thereby increasing the anion binding ability of the host. This assumption was further supported by a calculation of the electrostatic potential surfaces of **L** at the M06-2X/6-31G(d,p) level of theory (discussed later), showing the highest electron density on cyano groups and the most positive potential on the NH groups (Scheme 1). In addition, the conformational flexibility with six H-donor groups may allow the binding of a spherical halide within the ligand's cavity. We now report the results of halide binding studies of **L** in solution, structural aspects of several complexes, and computational studies. In particular, we show that the urea-based tripodal receptor has a significant selectivity for the fluoride anion in DMSO-$d_6$, which is further confirmed by 2D NOESY experiments.

## Experimental Section

### General

All the chemicals were purchased as reagent grade and were used without further purification. Nuclear magnetic resonance (NMR) spectra were recorded on a Varian Unity INOVA 500 FT-NMR. Chemical shifts for samples were measured in DMSO-$d_6$ and calibrated against sodium salt of 3-(trimethylsilyl) propionic-2,2,3,3,-$d_4$ acid (TSP) as an external reference in a sealed capillary tube. All NMR data were processed and analyzed with MestReNova Version 6.1.1-6384. IR spectra were recorded in KBr pellets on a Perkin Elmer FT-IR spectrometer. Mass spectral data were obtained in the ESI-MS positive mode on a FINNIGAN LCQDUO. The melting point was determined on a Mel-Temp (Electrothermal 120 VAC 50/60 Hz) melting point apparatus and was uncorrected. Elemental analysis was carried out by Columbia Analytical Services (Tucson, AZ 85714). All the structures reported here were analyzed from the X-ray laboratory at the University of Oklahoma.

**Synthesis**

**L**. Tris(2-aminoethyl) amine (300 mg, 2.05 mmol) was added with *p*-cyanophenyl isocyanate (886 mg, 6.15 mol) in chloroform at room temperature under constant stirring. The mixture was refluxed for 3 h. The precipitate was collected by filtration which was washed by chloroform and dried to give the neutral host. Yield: 1.07 g, 90%. $^1$H NMR (500 MHz, DMSO-$d_6$, TSP): δ 9.14 (s, 3H, Ar-N*H*), 7.67 (d, J = 8.5 Hz, 6H, Ar*H*), 7.58 (d, *J* = 8.5 Hz, 6H, Ar*H*), δ 6.39 (m, $J_1$ = 5.5 Hz, $J_2$ = 5.05 Hz, 3H, CH$_2$N*H*), 3.24 (m, $J_1$ = 6.1 Hz, $J_2$ = 5.95 Hz, $J_3$ = 5.90 Hz, 6H, NHC*H*$_2$), 2.64 (t, $J_1$ = 6.55 Hz, $J_2$ = 6.45 Hz, 6H, NC*H*$_2$). $^{13}$C NMR (125 MHz, DMSO-$d_6$,): δ 154.9 (*C*=O), 145.4 (Ar-*C*), 133.6 (Ar-*C*H), 120.0 (Ar-*C*N), 118.0 (Ar-*C*H), 102.8 (Ar*C*-CN)), 54.0 (NH*C*H$_2$), 37.9 (N*C*H$_2$), ESI-MS: m/z (+) 579.63 [M+H]$^+$. Anal. Calcd. for C$_{30}$H$_{30}$N$_{10}$O$_3$: C, 62.27; H, 5.23; N, 24.21. Found: C, 62.36; H, 5.24; N, 24.22. Crystals suitable for X-ray analysis were grown in two days from slow evaporation of the acetonitrile solvent at room temperature.

**[HL](Cl), 1**. The neutral host **L** (25 mg) was suspended in MeOH (10 mL), and a few drops of 40% HCl (approx. 5 drops) were added to the mixture. After stirring for 30 mins, the clear solution was kept at room temperature. Prism-shaped crystals suitable for X-ray analysis were grown from this solution by slow evaporation after one week. ESI-MS: m/z (+) 616.09 [M+H]$^+$. Anal. Calcd (%) for C$_{30}$H$_{31}$ClN$_{10}$O$_3$: C, 58.58; H, 5.07; N, 22.77. Found: C, 58.49; H, 5.05; N, 22.73.

**[HL](Br), 2**. The neutral host **L** (25 mg) was suspended in MeOH (10 mL), and a few drops of 49% HBr (approx. 5 drops) were added in the mixture. After stirring for 30 mins, the clear solution was kept at room temperature. Block-shaped crystals suitable for X-ray analysis were grown from this solution by slow evaporation after one week. ESI-MS: m/z (+) 559.53 [M+H]$^+$. Anal. Calcd (%) for C$_{30}$H$_{31}$BrN$_{10}$O$_3$: C, 54.63; H, 4.73; N, 21.23. Found: C, 54.69; H, 4.76; N, 21.21.

**[HL]$_2$(SiF$_6$)·6.35(H$_2$O), 3**. In an attempt to prepare the fluoride salt of **L**, silicon hexafluoride salt was obtained. The neutral host **L** (25 mg) was suspended in MeOH (10 mL), and a few drops of 49% HF (approx. 5 drops) were added in the mixture. After stirring for 30 mins, the clear solution was kept at room temperature. Prism-shaped crystals suitable for X-ray analysis were grown from

this solution by slow evaporation after three days. Due to an insufficient amount of crystals obtained, no characterization was carried out for this salt except X-ray crystallography.

**NMR studies**

$^1$H NMR titration studies were done to determine the binding constants of **L** for halides (F$^-$, Cl$^-$, Br$^-$ and I$^-$) in DMSO-$d_6$ at room temperature. [$n$-Bu$_4$N]$^+$X$^-$ was used as a source of the anion. Initial concentrations were [ligand]$_0$ = 2 mM, and [anion]$_0$ = 20 mM. Each titration was performed by 13 measurements at room temperature. The association constant $K$ was calculated by fitting of two NH signals with a 1:1 binding model, using the equation, $\Delta\delta = ([A]^0 + [L]^0 + 1/K - (([A]^0 + [L]^0 + 1/K)^2 - 4[L]^0[A]^0)^{1/2}) \Delta\delta_{max} / 2[L]^0$ (where **L** is the ligand and A is the anion).[46] The error limit in $K$ was less than 10%. $^{19}$F NMR studies were performed using 20 mM of [$n$-Bu$_4$N]$^+$F$^-$ in DMSO-$d_6$ at 25 °C. $^{19}$F NMR spectra were recorded for the fluoride solution before and after the addition of **L** (20 mM in DMSO-$d_6$), while a solution of NaF in D$_2$O in a sealed capillary tube was used as an external reference.

**DFT calculations**

In order to quantitatively understand the unique bonding within the tripodal urea ligand, density functional theory (DFT) calculations were carried out on each of the three F$^-$, Cl$^-$, and Br$^-$ anions. All quantum chemical calculations were carried out with the recent M06-2X meta-GGA hybrid functional, which has been shown to accurately predict the binding energies of ions and other noncovalent bonding interactions in large molecular systems.[47,48] Molecular geometries (including the empty ligand) were completely optimized without constraints at the M06-2X/6-31G(d,p) level of theory, and single-point energies with a very large 6-311+G(d,p) basis set were carried out in the presence of a polarizable continuum model (PCM) solvent model to approximate a DMSO environment (dielectric constant = 46.8).

**X-ray Crystallography**

The crystallographic data and details of data collection for the free ligand **L** and the anion complexes **1** - **3** are given in Table 1. Intensity data for all four samples were collected using a diffractometer with a Bruker APEX ccd area detector and graphite-monochromated MoKα radiation (λ = 0.71073 Å).[49,50] The samples were cooled to 100(2) K. The triclinic space group $P\bar{1}$ in **L** was determined by statistical tests and verified by subsequent refinement, while the monoclinic space group $P2_1/n$ in **1**, **2**, and **3** were determined by systematic absences and statistical tests with verification by subsequent refinement. The structure was solved by direct methods and refined by full-matrix least-squares methods on $F^2$.[51] The positions of hydrogens bonded to carbons were refined by a riding model. Hydrogens bonded to nitrogens were located on a difference map, and their positions were refined independently. Non-hydrogen atoms were refined with anisotropic displacement parameters. Hydrogen atom displacement parameters were set to 1.2 times the isotropic equivalent displacement parameters of the bonded atoms. Hydrogen-bonding interactions are shown in Table 2.

**Results and Discussion**

**Synthesis**

The synthesis of **L** was previously reported by our group and obtained as a pure product in high yield. A similar approach was also employed to synthesize related urea-based hosts by other groups.[38-40] Attempts to prepare complexes of neutral receptors with tetrabutyl ammonium halides were unsuccessful; therefore, **L** was converted to chloride and bromide salts by reacting with corresponding acids in methanol. The addition of hydrofluoric acid to the methanolic solution of **L** led to the formation of silicon hexafluoride ($SiF_6^{2-}$) salts due to the corrosion effect of HF to the glass vial.[34] The compound was fairly stable under acidic condition, allowing for the protonation at the tertiary amine. All the salts were characterized by single crystal structure analysis.

**NMR titrations**

The binding properties of **L** for halides were examined by $^1$H NMR titration studies using [$n$-Bu$_4$N]$^+$X$^-$ (X$^-$ = F$^-$, Cl$^-$, Br$^-$ and I$^-$) in DMSO-$d_6$. Figure 1 shows the $^1$H NMR spectra of the ligand before and after addition of one equivalent halide ion. In the $^1$H NMR of free **L**, two NH protons at the different chemical environment appear at 9.14 (H2) and 6.39 (H1) ppm. The addition of F$^-$ to **L** resulted in a significant downfield shift of both NH signals ($\Delta\delta$ =1.78 ppm for H2 and $\Delta\delta$ = 0.86 ppm for H1), suggesting an interaction of the anion with NH groups. A similar trend, although to a lesser extent, was observed for those protons upon the addition of Cl$^-$. However, in the case of Br$^-$ or I$^-$, there was little change in the chemical shifts. Figure 2a shows the stacking of $^1$H NMR titration spectra obtained from the experiments with portion-wise additions of chloride ion (0 to 10 equivalents), displaying a systematic shift change in the NH signals. The changes in the chemical shift of NH peaks of the ligand were recorded with an increasing amount of halide solution at room temperature, giving the best fit for a 1:1 binding model (Figure 2b). The 1:1 stoichiometry of the halide complex in solution was further verified by a Job plot displaying a maximum at an equimolar ratio of the anion and **L** (Figure S6). The binding constants of **L** were determined from a non-linear regression analysis of NH shift changes, showing a binding trend in the order of F$^-$ > Cl$^-$ > Br$^-$ > I$^-$. Specially, the ligand **L** showed a strong affinity for fluoride anion (log $K$ = 4.51) compared to the chloride anion (log $K$ = 3.09). This data suggests that the binding is largely dominated by the relative electronegativity and size of the anions. The highest binding for F$^-$ could be the results of the strong electrostatic interactions of this anion with the acidic NH of the host. The observed binding constant for fluoride is higher than that reported in a related host (log $K$=4.06).[40]

$^{19}$F NMR spectroscopy was also used to identify the chemical environment of the fluoride in the complex. Figure 3a shows the $^{19}$F NMR spectrum showing two peaks at -122.5 and -105.2 ppm for free [$n$-Bu$_4$N]$^+$F$^-$ (20 mM) in the presence of a NaF reference used in a sealed capillary tube in D$_2$O. The former peak is assigned to the reference fluoride ion solvated with D$_2$O, while the later is due to the fluoride ion of [$n$-Bu$_4$N]$^+$F$^-$ in DMSO-$d_6$. As clearly shown in Figure 3b-d, the addition of **L** (20 mM) to the fluoride solution resulted in a gradual downfield shift of the free fluoride resonance,

indicating the hydrogen bonding interactions of the fluoride ion with the NH groups of **L**. In particular, we observed a significant downfield shift of about 15 ppm after the addition of one equivalent of **L**. A similar trend in downfield shifts was previously reported for fluoride binding with an amide-based cryptand receptor.[52]

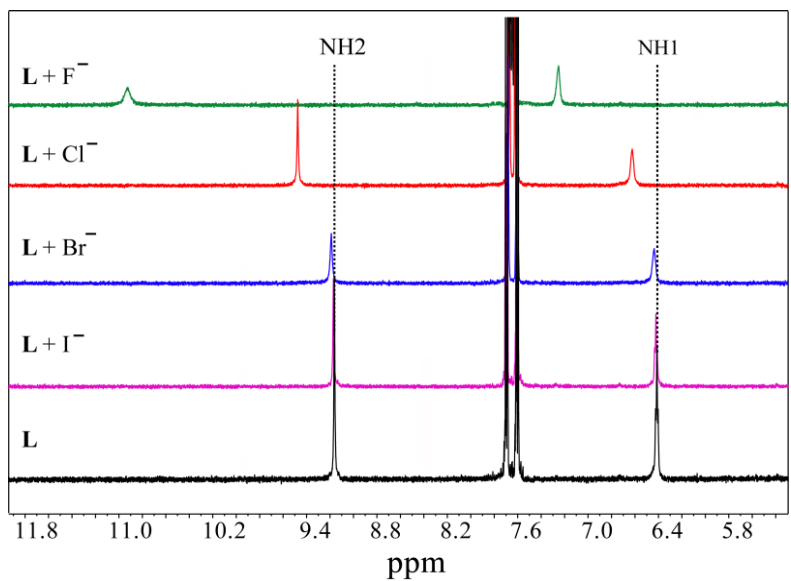

**Figure 1.** Partial $^1$H NMR spectra of **L** (2 mM) in the absence and presence of an anion showing two NH peaks in DMSO-$d_6$. An equivalent amount of [$n$-Bu$_4$N]$^+$X$^-$ was added to the ligand solution.

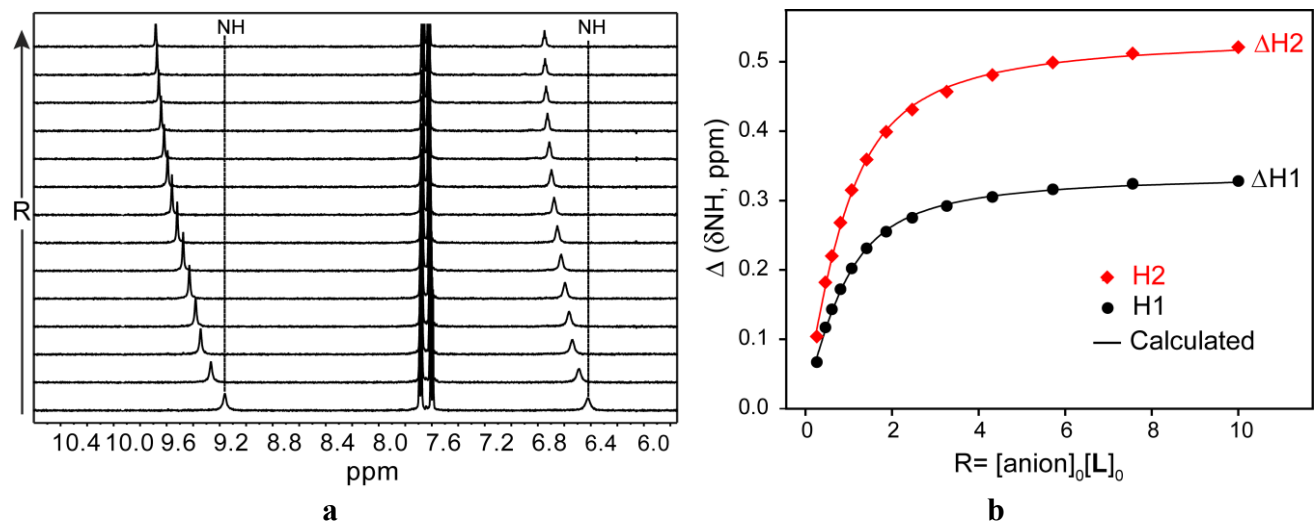

**Figure 2.** (a) Partial $^1$H NMR titration of **L** (2 mM) with an increasing addition of $n$-Bu$_4$N$^+$Cl$^-$ (20 mM) in DMSO-$d_6$, (b) $^1$H NMR titration curves of **L** with $n$-Bu$_4$N$^+$Cl$^-$ in DMSO-$d_6$ showing the net changes in the chemical shifts of NH (H1 = CH$_2$NHCO and H2 = CONHAr).

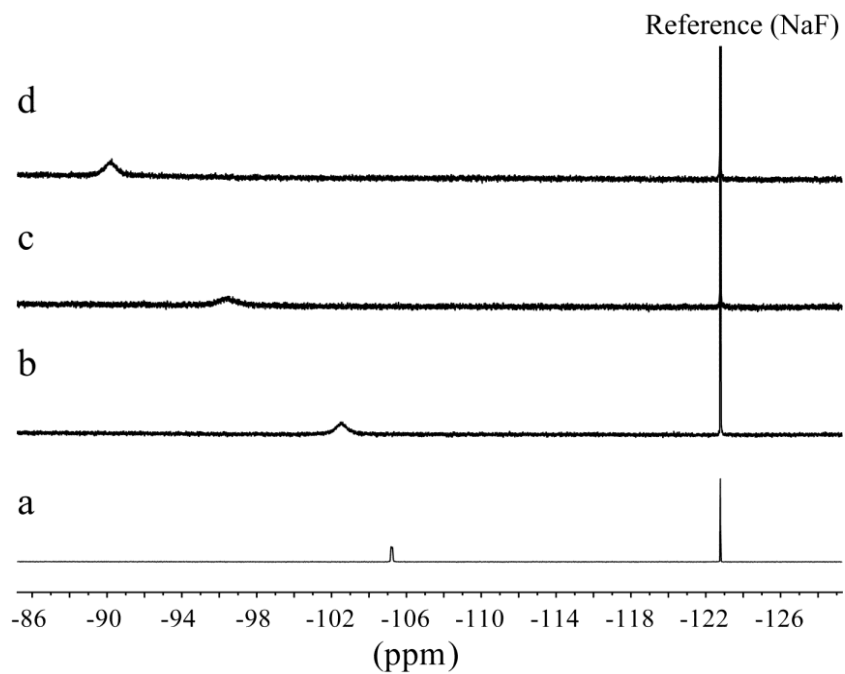

**Figure 3.** Partial $^{19}$F NMR spectra of $n$-Bu$_4$N$^+$F$^-$ in DMSO-$d_6$ at room temperature, showing the downfield shift of F$^-$ resonance upon the gradual addition of **L**. (a) Free $n$-Bu$_4$N$^+$F$^-$ ($\delta_F$ = -105.2 ppm), (b) $n$-Bu$_4$N$^+$F$^-$ + 0.25 equiv. of **L** ($\delta_F$ = -102.5 ppm), (c) $n$-Bu$_4$N$^+$F$^-$ + 0.50 equiv. of **L** ($\delta_F$ = -96.4 ppm), and (d) $n$-Bu$_4$N$^+$F$^-$ + 1.0 equiv. of **L** ($\delta_F$ = -90.1 ppm).

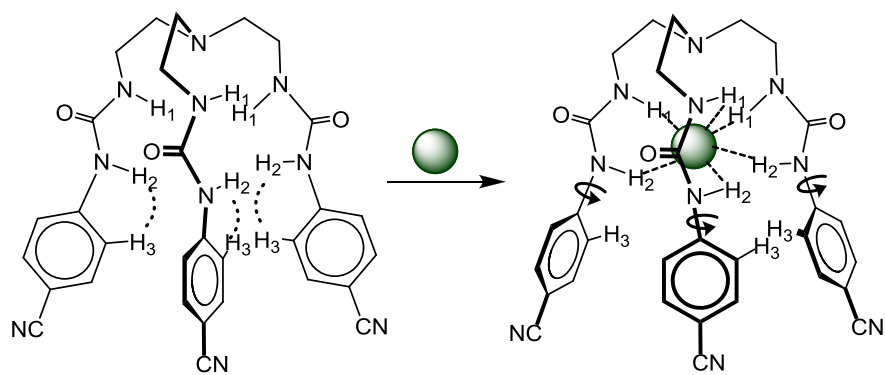

**Scheme 2.** Proposed binding mechanism of **L** for an anion in solution.

In order to characterize the solution structure of the complexes, 2D NOESY NMR experiments were carried out in DMSO-$d_6$ at room temperature. In the 2D NOESY NMR, the free ligand showed two strong NOESY contacts between H2···H3 and H1···H2 (Scheme 2 and Figure 4), which could be due to the fact that the aromatic plane connected to a urea unit is co-planar with the NH group. Such an assumption is further supported by the single crystal structure analysis of free ligand (discussed later). Upon addition of one equivalent of fluoride ion, all the NOESY contacts disappeared, indicating a conformational change of **L** due to the encapsulation of the anion. Indeed, the ligand showed high affinity for fluoride (log$K$ = 4.51) in DMSO-$d_6$, as discussed in the previous section. The encapsulation was further confirmed by molecular modeling studies performed in the same solvent environment (discussed later). Similar conformational changes were previously reported by Werner and Schneider in the optimized structure of chloride complex of a tren-based urea ligand.[53] The addition of one equivalent of chloride led to the disappearance of both NOE contacts (H2⋯H3 and H1⋯H2) in the 2D NOESY NMR spectra. However, the addition of bromide or iodide apparently did not affect the NOE contacts in the ligand, which could be due to the very weak interaction for this anion, as also supported by NMR titration data (Table 3). This observation further supports the formation of an encapsulated complex of **L** with fluoride or chloride ion in solution.

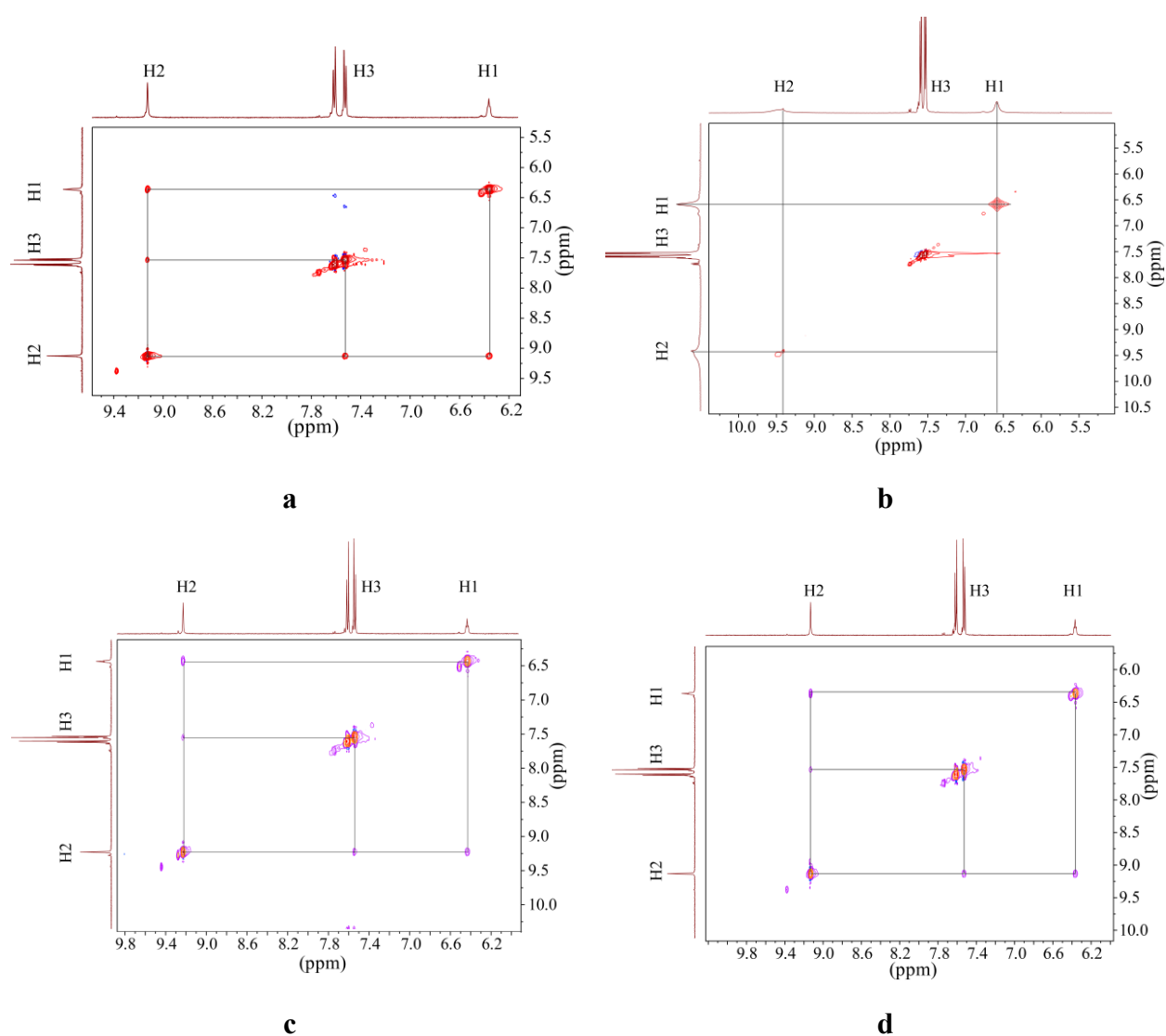

**Figure 4**. 2D NOESY NMR spectra of (a) free **L**, and **L** in the presence of one equivalent of (b) fluoride, (c) bromide, and (d) iodide anions in DMSO-$d_6$ at room temperature.

**DFT calculations**

The binding properties were also evaluated by DFT calculations for the free ligand and its halide complexes (except iodide, where the 6-311+G(d,p) basis set is not available) using a PCM model in a DMSO environment (dielectric constant = 46.8). In order to correlate binding strengths, the stabilization energy for each structure was calculated as $E_s$ = E(ligand) + E(anion) - E (ligand + ligand), and the binding energies for fluoride, chloride and bromide complexes were found to be 32.87, 12.90, and 8.49 kcal/mol, respectively. These values fairly correlate with the data of binding constants (Table 3), showing the highest binding affinity for fluoride. Figure 5 shows the optimized geometries of the fluoride and chloride anions bound to the neutral ligand. From the DFT-optimized geometries, we found that although six hydrogen bonds were formed with the individual anion, each of the conformations showed very distinct binding energies and configurations. Specifically, for the fluoride complex, the three arms were twisted, and the anion was tightly bound inside the cavity, with the NH···F bond distances of 2.689 to 2.907 Å (Figure 5a). The coordination patterns and bond distances are comparable with the structure of fluoride complex with a pentafluorophenyl-substituted tripodal urea (NH···F = 2.700(3) to 2.884(3) Å) reported by Ghosh et al.[40] As shown in Table 4, the average bond NH···X distances in the optimized geometries are 2.78, 3.36, and 3.45 Å for the fluoride, chloride, and bromide complexes, respectively. The NH···X distances in the optimized structures of [**L**(Cl)] and [**L**(Br)] are also close to the corresponding values obtained from the crystal structures ranging from 3.1802(18) to 3.5679(18) Å for [H**L**](Cl) and 3.335(3) to 3.645(3) Å for [H**L**](Br). In particular, we obtained a considerably high binding energy for the fluoride complex compared to the chloride or bromide complex, which could be due to the high electron density of the fluoride anion, making it a stronger H-bond acceptor. In the case of chloride or bromide complex, each was shown to form an almost perfect $C_3$ symmetric complex (for the chloride complex, see Figure 5b), although the bromide anion was loosely held due its larger size and lower charge density. This observation also agrees with the NOESY results, showing the disappearance of certain NOE contacts for fluoride complex (Figure 4b).

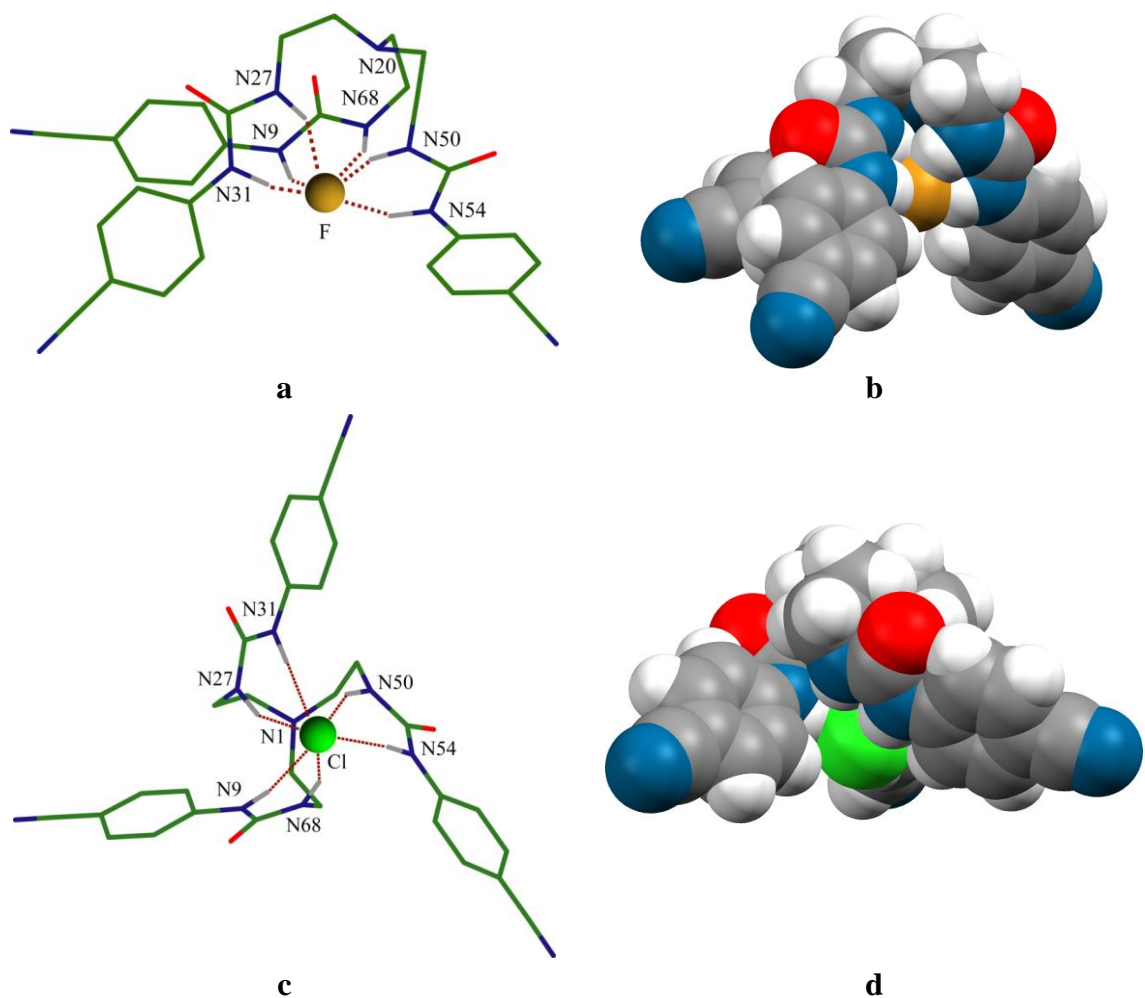

**Figure 5**. Optimized structures: (a) perspective view and (b) space filling model of the fluoride complex of **L**; and (c) perspective view and (d) space filling model of the chloride complex of **L**.

**Crystal Structure analysis**

The free ligand **L** crystallizes in the triclinic $P\bar{1}$ space group. The structural analysis of **L** shows that it forms a pseudo-cavity with three arms suitable for hosting an anion (Figure 6a). The cavity possesses an approximate $C_3$ symmetry axis passing through the tertiary N atom. Two aromatic units connected to N21 and N35 are involved in CH···π interactions with a centroid···centroid distance of 3.652 Å, with two nitrogens N18 and N32 (3.715 Å) in close proximity. Similar interactions were reported before for a thiophene-based tripodal amine.[54] Each urea unit has a usual *anti-anti* conformation with respect to NH and the carbonyl O. The aromatic planes connected to the urea units are almost co-planar with the NH groups, as indicated by torsion angles close to 180°. Two NH groups (N4 and N7) in one arm form two strong intramolecular hydrogen bonds with one carbonyl oxygen (O20) of another arm. The H-bond distances are N4···O20 = 2.9397(18) and N7···O20 = 2.8849(17) Å. The molecule forms a centrosymmetric dimer from the interactions of four intermolecular H-bonds (Figure 6b). As shown in Figure 6c, two units in the dimer are antiparallel to each other.

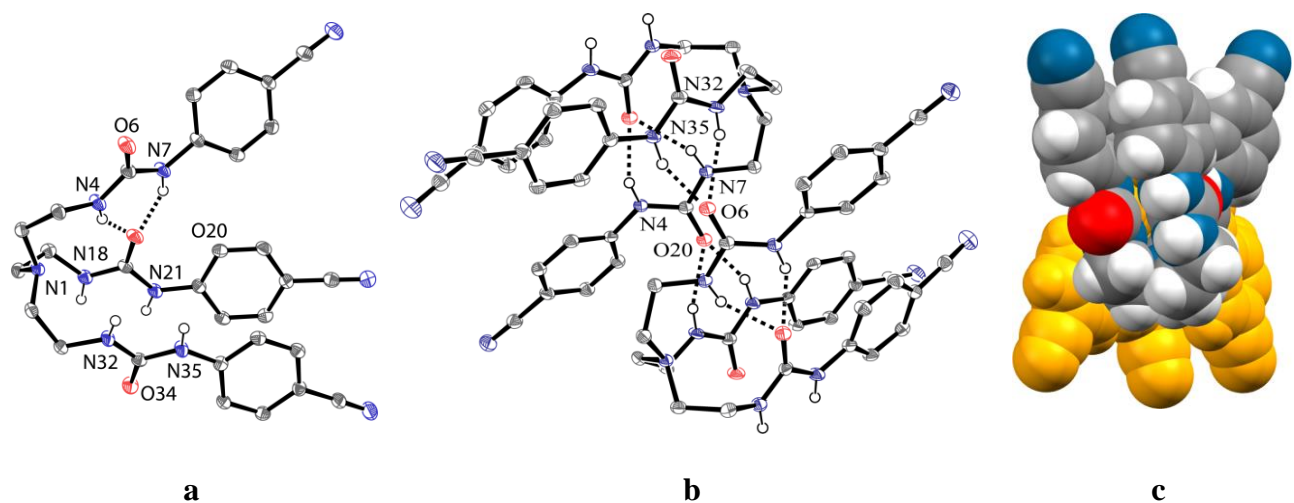

**Figure 6**. ORTEP drawing of (a) free **L** and (b) its dimer, with thermal ellipsoids at the 50% probability level (hydrogen atoms on carbons are omitted for clarity); (c) space filling view of dimeric **L**.

The chloride salt of the ligand prepared from the reaction of **L** with HCl in ethanol crystallizes as [H**L**.(Cl)] in the monoclinic space group *P*2$_1$/*n*. The tertiary amine is protonated as expected, and the charge is balanced by one chloride ion. The proton on the tertiary amine (N1) points inside the tripodal cavity and is held by a strong H-bond with one endo-oriented carbonyl oxygen (O34) of a urea group, with a distance of N···O = 2.744(2) Å. Another intermolecular hydrogen bonding interaction is observed between N18 of one urea and O6 of other urea with a N···O distance of 2.912(2) Å. Therefore, the cavity is apparently not favorable for accommodating a chloride in the solid state (Figure 7a). As a result, the anion remains outside the cavity bonded to one urea unit with two strong hydrogen bonds (N4···Cl = 3.2776(18) and N7···Cl = 3.1802(18) Å). This chloride is further coordinated with a neighboring tren via one strong H-bond (N35···Cl = 3.2016(18) Å) and one relatively weak H-bond (N32···Cl = 3.5679(18) Å), resulting in the formation of a centrosymmetric dimer (Figures 7b and 7c). Therefore, the chloride ion is coordinated with a total of four bonds in a tetragonal pyramidal fashion where the anion is located on the vertex of pyramid. Hydrogen bonding interactions with the coordinating chloride ion is shown in Figure 7b. In the crystal, all NH groups except N21 are involved as H-bond donors either for the chloride or carbonyl oxygen. As viewed in the packing diagram along the *c* axis (Figure 7d), the molecules are assembled to generate a rod-like structure through NH ···Cl interactions (along the *b* axes) and several short contacts between CN (cyano) and CH (aliphatic) groups (along the *a* axes).

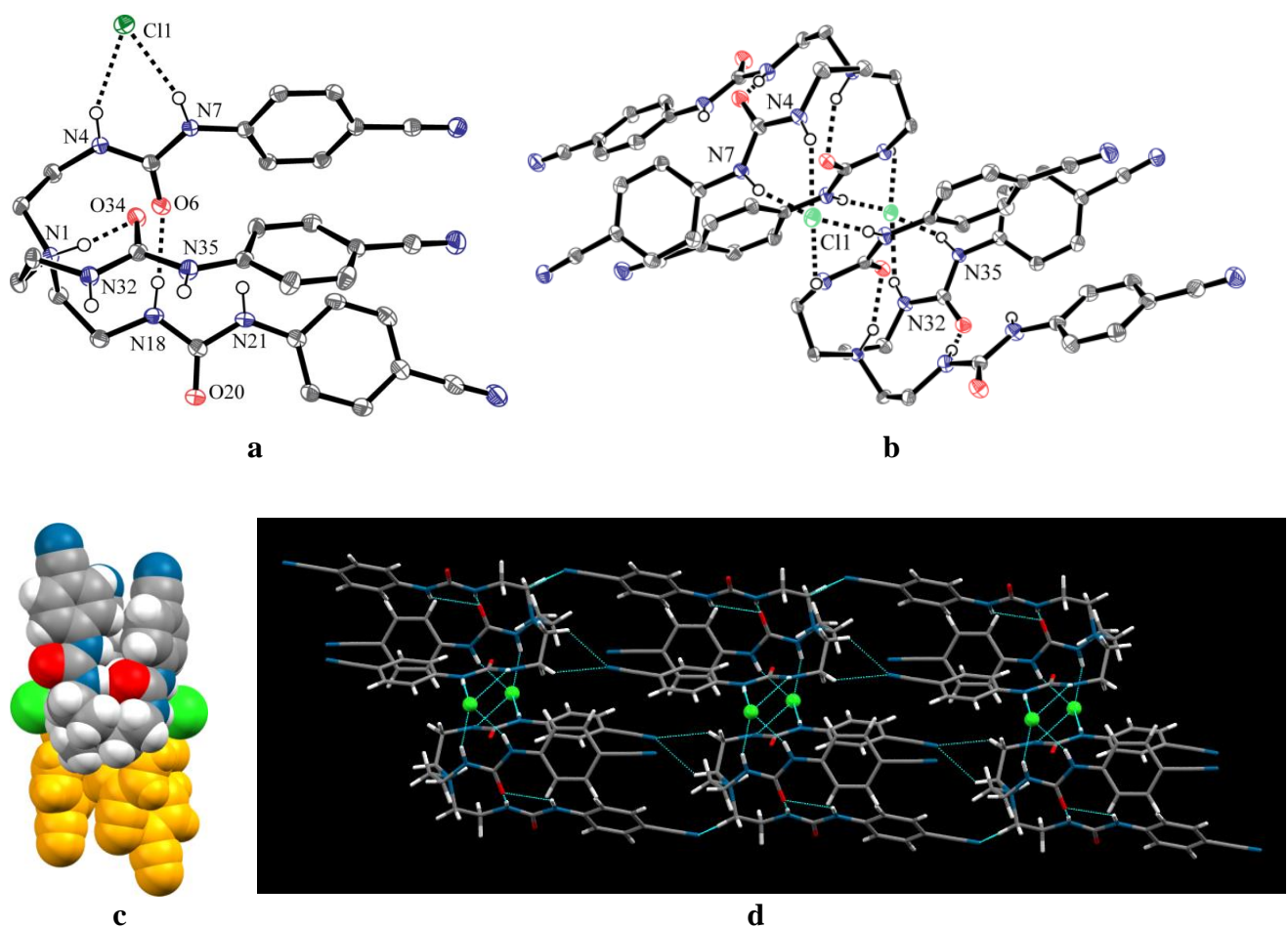

**Figure 7**. ORTEP drawing of (a) [H**L**](Cl) and (b) its dimer, with thermal ellipsoids at the 50% probability level (hydrogen atoms on carbons are omitted for clarity); (c) space filling view of dimeric [H**L**](Cl); (d) lattice structure of [H**L**](Cl) viewed along the along *c* axis.

The structural aspects of bromide complex of receptor **L** are strikingly similar to that of the chloride complex. As shown in Figure 8a, the host binds a single bromide outside the cavity. The tertiary amine is found to be protonated, which points inside the cavity, making a hydrogen bond with one carbonyl oxygen (N1···O20 = 2.755(4) Å). The bromide is bonded to both NH groups of the single urea unit (N4···Br = 3.405(3) and N7···Br = 3.335(3) Å). Each urea unit with respect to the NH and carbonyl O is essentially planar. Two NH groups (N32 and N35) from a single urea unit are directed toward the cavity, while the remaining four NH groups are directed outside the cavity, serving as H-bond donors for externally-located bromide ions in a lattice. The details of the hydrogen bonding interactions are listed in Table 2. As shown in Figures 8b and 8c, two anti-parallel tripodal units are paired via two anions from opposite sites to form a dimer. Figure 8d shows the molecules are also packed with hydrogen bonding interactions and CN···CH short contacts to form a rod-like structure, in a similar fashion observed in the chloride complex.

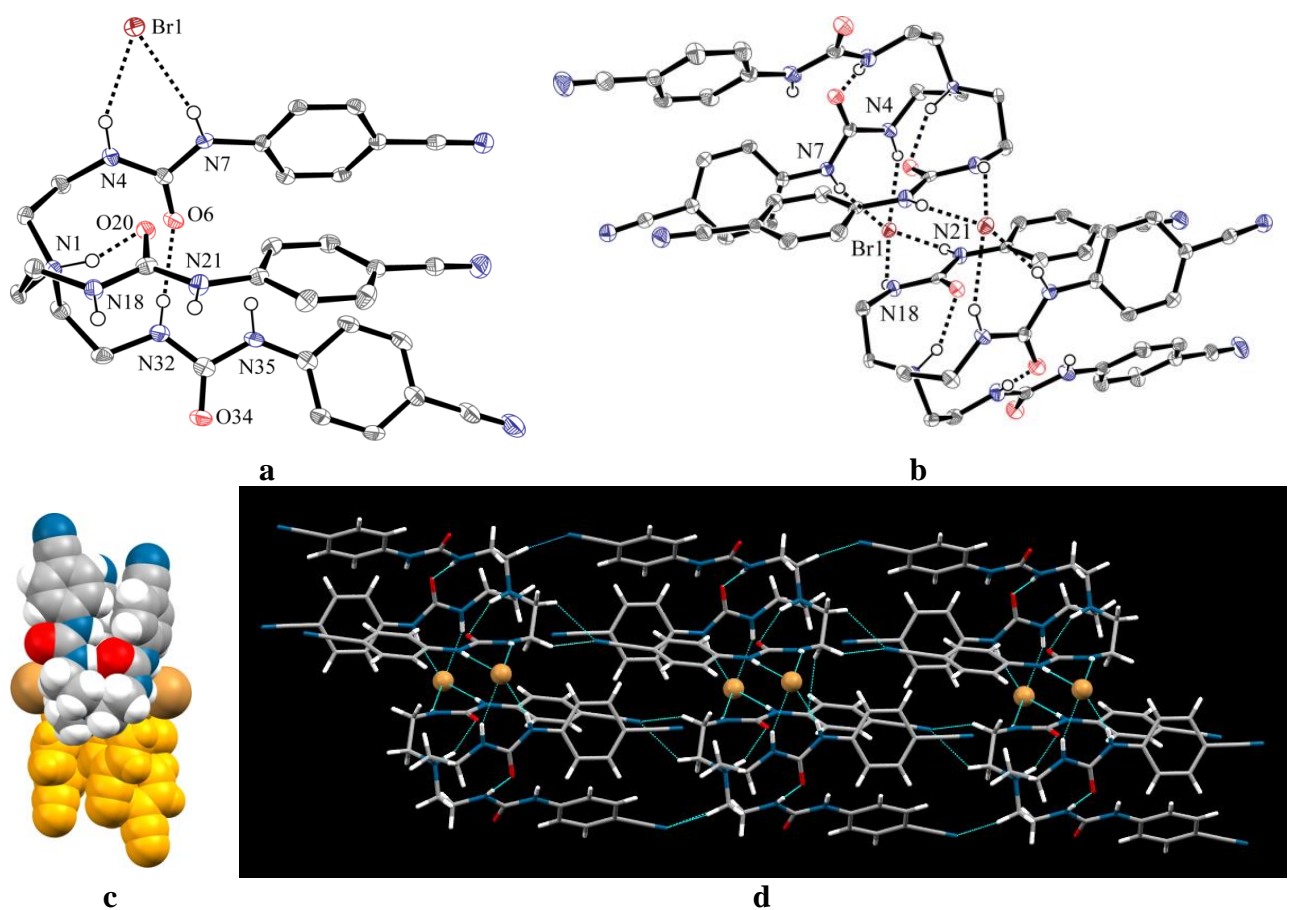

**Figure 8**. ORTEP drawing of (a) [HL](Br) and (b) its dimer, with thermal ellipsoids at the 50% probability level (hydrogen atoms on carbons are omitted for clarity); (c) space filling view of dimeric [HL](Br); (d) lattice structure of [HL](Br) viewed along the along *c* axis.

In an attempt to prepare the fluoride complex of **L**, hexafluorosilicate salt **3** was obtained from reaction of the host in methanol with HF acid in a glass vial. Obviously, the source of hexafluorosilicate in the system is due to the reaction of HF with the glass. The structure of this complex is included in this paper because of the interesting bonding aspects of $SiF_6^{2-}$ through F atoms with **L** and water. The X-ray analysis of this complex suggests that the salt crystallizes as [H**L**]$_2$(SiF$_6$)·6.35(H$_2$O) in the monoclinic space group $P2_1/n$. One water molecule that is directly bonded with $SiF_6^{2-}$ was ordered. Other water molecules were disordered and were modeled only with isotropic oxygen atoms.

The asymmetric unit contains two tripodal ureas where each urea is protonated at the terminal nitrogen. Therefore, the total charges are balanced by one di-negatively charged hexafluorosilicate ion. The intramolecular hydrogen bonding interactions are different than those observed in the chloride or bromide complex. In this case, the proton is internally bonded with two carbonyl O atoms with N···O distances of 2.834(2) and 3.028(2) Å, in contrast to the one carbonyl O atom in complex **1** or **2**. The coordination environment of the $SiF_6^{2-}$ is quite different than that observed in the chloride or bromide complex. In an asymmetric unit, the anion is held between two parallel tren units and one water molecule contributing six H-bonds (five NH···F bonds with urea groups and one OH···F bond with water molecule). The ORTEP view of the crystal structure is depicted in Figure 9a. Tasker *et al.* reported a complex of $PtCl_6^{2-}$ with two protonated tripodal tris-urea substituted with butyl groups, showing the participation of two arms from each receptor to form a sandwich type complex.[41] However, in our case, the anion is H-bonded with two arms from one receptor and one arm from other receptor.

As listed in Table 2, the NH···F bond distances are in the range from 2.823 to 3.349 Å. The complete coordination environment of $SiF_6^{2-}$ is shown in Figure 9b, where the anion is held by a total of 12 H-bonds and entrapped within a hole generated by six ligands and two water molecules (Figure 10). In the structure, the anion sits on a crystallographic center of symmetry. Among the six ligands, four are directly bonded to the central anion, while the remaining two are connected to the anion through water molecules.

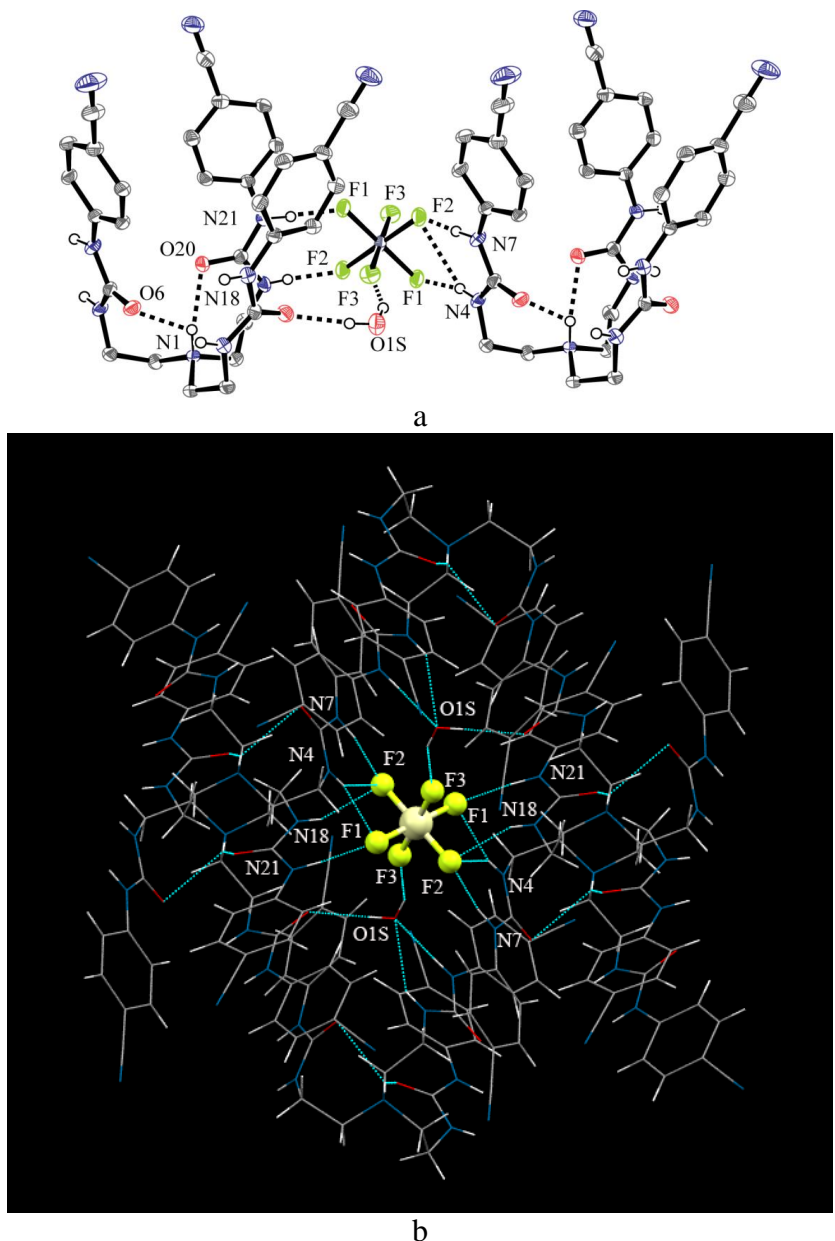

**Figure 9**. (a) ORTEP drawing of [**HL**]$_2$(SiF$_6$) with thermal ellipsoids at the 50% probability level (hydrogen atoms on carbons are omitted for clarity); (b) coordination environment of SiF$_6^{2-}$ showing a total of 12 H-bonds.

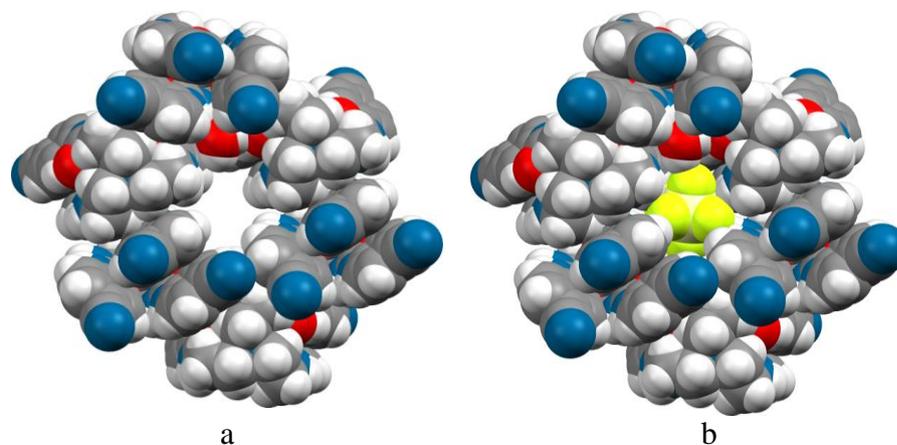

**Figure 10**. Space filling views of hexafluorosilicate comlex of **L**, (a) showing a hole generated by six ligands and two water molecules and (b) encapsulated $SiF_6^{2-}$.

**Conclusions**

We have presented a comprehensive study of a urea based tripodal receptor for spherical halide ions using both experimental and theoretical techniques. In particular, the receptor was shown to bind a fluoride ion strongly in solution compared to the other halide ions. The experimental observation from solution studies are clearly correlated with predictive methods from DFT calculations, indicating the formation of an encapsulated complex with hydrogen bond donors from NH groups. The binding constants in the order of fluoride > chloride > bromide > iodide, suggest that the binding is primarily dominated by the relative basicity of halides, which are also in line with the Hofmeister effect.[55] An important aspect in this study is the use of 2D NOESY spectroscopic techniques to characterize the solution structures. Additionally, $^{19}F$ NMR spectroscopy has been used to probe the chemical environment of fluoride in solution. Structural characterization of the chloride and bromide complexes grown in acidic medium suggests that the one halide ion is externally bonded with two receptors with four NH-bonds in both cases, where the tertiary nitrogen is protonated and points towards the cavity. The obvious discrepancy in solution and solid-state results could be due to the proton on the tertiary nitrogen, preventing the encapsulation of an anion in the cavity. Interestingly, the protonated receptors are assembled with water molecules to form a perfect cage to encapsulate a silicon hexafluoride anion. Since the report on selective neutral receptors is still in its infancy, the present solution and solid-state findings coupled with theoretical results further expand the understanding of binding mechanisms in host-guest complexes.

**Acknowledgments**

The National Science Foundation is acknowledged for a CAREER award (CHE-1056927) to MAH. This work was supported by the National Institutes of Health (G12RR013459). The 500 NMR instrument used for this work was funded by the National Science Foundation (CHE-0821357). The authors thank the National Science Foundation (CHE-0130835) and the University of Oklahoma for funds to acquire the diffractometer used in this work.

**Supporting Information Available**: Crystallographic data in CIF format and NMR and mass spectra, Job's plot, optimized structures and Cartesian coordinates for all optimized structures in PDF formats. This material is available free of charge via the Internet at http://pubs.acs.org.

**Table 1.** Crystallographic Data for **L**, [**HL**](Cl), **1**, [**HL**](Br), **2** and [**HL**]$_2$(SiF$_6$)·6.35(H$_2$O), **3**.

|  | L | 1 | 2 | 3 |
|---|---|---|---|---|
| Chemical formula | C$_{30}$H$_{30}$N$_{10}$O$_3$ | C$_{30}$H$_{31}$ClN$_{10}$O$_3$ | C$_{30}$H$_{31}$BrN$_{10}$O$_3$ | C$_{60}$H$_{74.70}$F$_6$N$_{20}$O$_{12.35}$Si |
| M | 578.64 | 615.10 | 659.56 | 1415.79 |
| Crystal system | Triclinic | monoclinic | monoclinic | monoclinic |
| $a$/Å | 8.7312(11) | 13.2509(18) | 13.2935(19) | 15.4907(14) |
| $b$/Å | 12.8400(17) | 11.3650(16) | 11.3515(17) | 11.5343(11) |
| $c$/Å | 13.6820(18) | 19.471(2) | 19.928(3) | 19.4685(17) |
| $\alpha$/° | 91.989(3) | 90.00 | 90.00 | 90.00 |
| $\beta$/° | 107.888(2) | 97.470(8) | 97.895(3) | 96.061(2) |
| $\gamma$/° | 100.753(2) | 90.00 | 90.00 | 90.00 |
| $V$/Å$^3$ | 1427.1(3) | 2907.4(6) | 2978.7(8) | 3459.1(5) |
| $T$/K | 100(2) | 100(2) | 100(2) | 100(2) |
| Space group | $P\bar{1}$ | $P2_1/n$ | $P2_1/n$ | $P2_1/n$ |
| Z | 2 | 4 | 4 | 2 |
| $\mu$/mm-1 | 0.092 | 0.184 | 1.431 | 0.124 |
| reflns measured | 16034 | 35737 | 28725 | 35178 |
| indept reflns | 7004 | 7232 | 6091 | 8552 |
| $R_{int}$ | 0.0320 | 0.0721 | 0.1087 | 0.0650 |
| [a] $R1$ (I > 2σ(I)) | 0.0495 | 0.0519 | 0.0540 | 0.0600 |
| w$R$(F2) values (I > 2σ(I)) | 0.1166 | 0.1083 | 0.1037 | 0.1408 |
| [b] w$R2$ (all data) | 0.1273 | 0.1202 | 0.1150 | 0.1581 |
| Goodness of fit | 1.003 | 1.010 | 1.012 | 1.003 |

[a] $R1 = \Sigma ||F_o| - |F_c|| / \Sigma |F_o|$, [a] $wR2 = \{ \Sigma [w(F_o^2 - F_c^2)^2] / \Sigma [w(F_o^2)^2] \}^{1/2}$

**Table 2.** Selected hydrogen bonding parameters (Å,°) for the ligand and different anion complexes.

| L | | | 1 | | |
|---|---|---|---|---|---|
| D–H···A | D···A | ∠D–H···A | D–H···A | D···A | ∠D–H···A |
| N(4)-H(4)...O(20) | 2.9397(18) | 154.0(17) | N(1)-H(1)...O(34) | 2.744(2) | 156.3(19) |
| N(7)-H(7)...O(20) | 2.8849(17) | 153.2(16) | N(4)-H(4)...Cl(1) | 3.2776(18) | 157.3(18) |
| N(18)-H(18)...O(34)[i] | 2.8925(17) | 160.6(16) | N(7)-H(7)...Cl(1) | 3.1802(18) | 162.0(18) |
| N(21)-H(21)...O(34)[i] | 3.0807(18) | 149.8(16) | N(18)-H(18)...O(6) | 2.912(2) | 176(2) |
| N(32)-H(32)...O(6)[ii] | 2.9315(18) | 150.2(16) | N(32)-H(32)...O(20)[iii] | 3.024(2) | 127(2) |
| N(32)-H(32)...O(6)[ii] | 2.9315(18) | 150.2(16) | N(32)-H(32)...Cl(1)[iv] | 3.5679(18) | 135.4(19) |
| | | | N(35)-H(35)...Cl(1)[iv] | 3.2016(18) | 175(2) |

| 2 | | | 3 | | |
|---|---|---|---|---|---|
| D–H···A | D···A | ∠D–H···A | D–H···A | D···A | ∠D–H···A |
| N(1)-H(1)...O(20) | 2.755(4) | 153(3) | N(1)-H(1)...O(20) | 2.834(2) | 143(2) |
| N(1)-H(1)...N(32) | 3.056(4) | 106(2) | N(1)-H(1)...O(6) | 3.028(2) | 122.7(19) |
| N(4)-H(4)...Br(1) | 3.405(3) | 157(3) | N(4)-H(4)...F(1)vi | 2.953(2) | 159(2) |
| N(7)-H(7)...Br(1) | 3.335(3) | 166(3) | N(4)-H(4)...F(2) | 3.266(2) | 142(2) |
| N(18)-H(18)...Br(1)[v] | 3.645(3) | 149(3) | N(7)-H(7)...F(2) | 2.833(2) | 171(2) |
| N(32)-H(32)...O(6) | 2.919(4) | 178(4) | N(18)-H(18)...F(2)[vii] | 2.978(2) | 156(3) |
| | | | N(18)-H(18)...F(3)[viii] | 3.349(3) | 146(2) |
| | | | N(21)-H(21)...F(1)[ix] | 2.823(2) | 173(2) |
| | | | (1S)-H(1SB)...F(3)[viii] | 2.842(2) | 156(3) |

Symmetry codes: (i) 1 -x, -y+1, -z; (ii) 2 -x+1, -y+1, -z+1; (iii) 1 -x+1, -y, -z; (iv) 2 -x+1, -y+1, -z; (v) 1 -x+1, -y+1, -z+1; (vi) 1 -x+1, -y, -z+1; (vii) 2 -x+1, -y+1, -z+1; (viii) x, y+1, z; (ix) -x+3/2, y-1/2, -z+3/2.

**Table 3.** Binding constants (log $K$) of tripodal *tris*-urea receptor (**L**) with halides in DMSO-$d_6$.

| Anion | log $K$ |
|---|---|
| Fluoride | 4.51 |
| Chloride | 3.09 |
| Bromide | 1.71 |
| Iodide | 1.01 |

**Table 4.** Selected hydrogen bonding parameters (A,°) for the halide complexes in their optimized strictures.

| | [**L**(F)]⁻ | | **L**(Cl)]⁻ | | **L**(Br)]⁻ | |
|---|---|---|---|---|---|---|
| D–H···A | D···A | ∠D–H···A | D···A | ∠D–H···A | D···A | ∠D–H···A |
| N(54)–H···X | 2.691 | 152.98 | 3.212 | 160.87 | 3.318 | 161.26 |
| N(50)–H···X | 2.802 | 148.94 | 3.465 | 150.48 | 3.613 | 151.15 |
| N(68)–H···X | 2.817 | 151.74 | 3.553 | 148.70 | 3.589 | 151.53 |
| N(9)–H···X | 2.764 | 151.80 | 3.220 | 160.80 | 3.334 | 164.96 |
| N((27)–H···X | 2.907 | 146.36 | 3.486 | 149.90 | 3.525 | 154.05 |
| N(31)–H···X | 2.689 | 160.78 | 3.228 | 162.76 | 3.329 | 158.47 |

**References:**

1. Bianchi, A. Bowman-James, K. García-España, E. *Supramolecular chemistry of anions*; Wiley-VCH: New York: 1997.

2. Gale, P. A.; Gunnlaugsson, T. *Chem. Soc. Rev.* **2010**, *39*, 3595–3596.

3. Wenzel, M.; Hiscock, J. R.; Gale, P. A. *Chem. Soc. Rev.* **2012**, *41*, 480–520.

4. Park, C. H.; Simmons, H. E. *J. Am. Chem. Soc.* **1968**, *90*, 2431–2433.

5.  Bell, R. A.; Christoph, G. G.; Fronczek, F. R.; Marsh, R. E. *Science* **1975**, *190*, 151–152.

6.  Llinares, J. M.; Powell, D.; Bowman-James, K. *Coord. Chem. Rev.* **2003**, *240*, 57–75.

7.  Bazzicalupi, C.; Bencini, A.; Bianchi, A.; Borsari, L.; Danesi, A.; Giorgi, C.; Mariani, P.; Pina, F.; Santarellia, S.; Valtancolia, B. *Dalton Trans.* **2006**, 5743–5752.

8.  Bondy, C. R.; Loeb, S. J. *Coord. Chem. Rev.* **2003**, *240*, 77–99.

9.  Hossain, M. A.; Llinares, J. M.; Powell, D.; Bowman-James, K. *Inorg. Chem.* **2001**, *40*, 2936–2937.

10. Kang, S. O.; Day, V. W.; Bowman-James, K. *Org. Lett.* **2009**, *11*, 3654–3657.

11. Begum, R.; Kang, S. O.; Bowman-James, K. *Angew. Chem. Int. Ed.* **2006**, *45*, 7882–7894.

12. Hossain, M. A.; Kang, S. K.; Llinares, L. M.; Powell, D.; Bowman-James, K. *Inorg. Chem.* **2003**, *42*, 5043–5045.

13. Inoue, Y.; Kanbara, T.; Yamamoto, T. Thioamide. *Tetrahedron Lett.* **2003**, *44*, 5167–5169.

14. Amendola, V, Fabbrizzi, L.; Mosca, L. *Chem. Soc. Rev.* **2010**, *39*, 3889–3915.

15. Custelcean, R. *Chem. Commun.* **2008**, 295–307.

16. Carroll, C. N.; Berryman, O. B.; Johnson II, C. A.; Zakharov, L. N.; Haley, M. M; Johnson, D. W. *Chem. Commun.* **2009**, 2520–2522.

17. Zhang, Z.; Schreiner, P. R. *Chem. Soc. Rev.* **2009**, *38*, 1187–1198.

18. Li, A-F.; Wang, J-H.; Wang, F.; Jiang, Y-B. *Chem. Soc. Rev.* **2010**, *39*, 3729–3745.

19. Sessler J. L.; Camiolo S.; Gale P .A. *Coord. Chem. Rev*. **2003**, *240*, 17–55.

20. Custelcean, R.; Delmau, L. H.; Moyer, B. A.; Sessler, J. L.; Cho, W.-S.; Gross, D.; Bates, G. W.; Brooks, S. J.; Light, M. E.; Gale, P. A. *Angew. Chem. Int. Ed.* **2005**, *44*, 2537–2542.

21. Sessler, J. L.; Gross, D. E.; Cho, W.-S.; Lynch, V. M; Schmidtchen, F. P.; Bates, G. W.; Light, M. E.; Gale. P. A. *J. Am. Chem. Soc.* **2006**, *128*, 12281–12288.

22. Bates, G. W.; Triyanti, L. M. E.; Albrecht, M.; Gale, P. A. *J. Org. Chem.* **2007**, *72*, 8921–8927.

23. Chang, K.-J.; Moon, D.; Lah, M. S.; Jeong, K.-S. *Angew. Chem. Int. Ed.* **2005**, *44*, 7926–7929.

24. Chang, K.-J.; Chae, M.-K.; Lee, C.; Lee, J.-Y.; Jeong, K.-S. *Tetrahedron Lett.* **2006**, *47*, 6385–6388.

25. Caltagirone, C.; Gale, P. A.; Hiscock, J. R.; Brooks, S. J.; Hursthouse, M. B.; Light, M. E. *Chem. Commun.* **2008**, 3007–3009.

26. Fan, E.; Van Arman, S. A.; Kincaid, S.; Hamilton, A. D. *J. Am. Chem. Soc.* **1993**, *115*, 369–370.

27. Boiocchi, M.; Del Boca, L.; Esteban-Gomez, D.; Fabbrizzi, L., Licchelli, M., Monzani, E. *J. Am. Chem. Soc.* **2004**, *126*, 16507–16514.

28. Albrecht, M.; Triyanti, Schiffers, S.; Osetska, O.; Raabe, G.; Wieland, T.; Russo, L.; Rissanen, K. *Eur. J. Org. Chem.* **2007**, *17*, 2850–2858.

29. Bates, G. W.; Gale, P. A.; Light, M. E. *Chem. Commun.* **2007**, 2121–2123.

30. Johnson, C. A.; Berryman, O. B.; Sather, A. C.; Zakharov, L. N.; Haley, M. M.; Johnson, D. W. *Cryst. Growth Des*. **2009**, *9*, 4247–4249.

31. Ros-Lis, J. V.; Martinez-Máñez, R.; Sancenon, F.; Soto, J.; Rurack, K.; Weißhoff, H. *Eur. J. Org. Chem.* **2007**, *15*, 2449–2458.

32. McKee, V.; Nelson, J.; Town, R. M. *Chem. Soc. Rev.* **2003**, *32*, 309–325.

33. Hossain, M. A. *Curr. Org. Chem*. **2008**, *12*, 1231–1256.

34. Hossain, M.A.; Morehouse, P.; Powell, P. D.; Bowman-James. K. *Inorg. Chem*. **2005**, *44*, 2143–2149.

35. Lakshminarayanan, P. S.; Kumar, D. K.; Ghosh, P. *Inorg. Chem*. **2005**, *44*, 7540–7546.

36. Hossain, M. A.; Saeed, M. A.; Gryn'ova, G.; Powell, D. R.; Leszczynski, J. *CrystEngComm.* **2010**, *12,* 4042– 4044.

37. Saeed, M. A.; Fronczek, F. R.; Huang, M. J. Hossain, M. A. *Chem. Commun*. **2010**, *46*, 404–406.

38. Custelcean, R.; Moyer, B. A.; Hay, B. P. *Chem. Commun.* **2005**, 5971–5973.

39. Wu, B.; Liang, J.; Yang, J.; Jia, C.; Yang, X.-J.; Zhang, H.; Tangb, N; Janiak, C. *Chem. Commun.* **2008**, 1762–1764.

40. Ravikumar, I.; Lakshminarayanan, P. S.; Arunachalam, M.; Suresh, E.; Ghosh, P. *Dalton Trans*. **2009**, 4160–4168.

41. Bell, K. J.; Westra, A. N.; Warr, R. J.; Chartres, J.; Ellis, R.; Tong, C. C.; Blake, A. J.; Tasker, P. A.; Schroder, M. *Angew. Chem. Int. Ed.* **2008**, *47*, 1745–1748.

42. Custelcean, R.; Bock, A.; Moyer, B. A. *J. Am. Chem. Soc.* **2010**, *132*, 7177–7185.

43. Li, M.; Wu, B; Cui, F.; Hao, Y.; Huang, X.; Yang, X.-J. *Anorg. Allg. Chem.* **2011**, *637,* 2306–2311.

44. Busschaert, N.; Wenzel, M.; Light, M. E.; Iglesias-Hernandez, P.; Perez-Tomas, R.; Gale, P. A. *J. Am. Chem. Soc.* **2011**, *133*, 14136–14148.

45. Pramanik, A.; Thompson, B.; Hayes, T.; Tucker, K.; Powell, D. R.; Bonnesen, P. V.; Ellis, E. D.; Lee, K. S.; Yua, H.; Hossain, M. A. *Org. Biomol. Chem.* **2011**, *9*, 4444–4447.

46. Schneider, H. J.; Kramer, R.; Simova, S.; Schneider, U. *J. Am. Chem. Soc.* **1988**, *110*, 6442–6448.

47. Truhlar, D. G. *J. Chem. Theory Comput.* **2006**, *2*, 364–382.

48. Truhlar, D. G. *J. Chem. Theory Comput.* **2008**, *4*, 1849–1868.

49. Data Collection: SMART Software Reference Manual, **1998**. Bruker-AXS, 5465 E. Cheryl Parkway, Madison, WI 53711–5373, USA.

50. Data Reduction: SAINT Software Reference Manual, **1998**. Bruker-AXS, 5465 E. Cheryl Parkway, Madison, WI 53711–5373, USA.

51. Sheldrick, G. M. *Acta Cryst*. **2008**, *A64*, 112–122.

52. Kang, S. O.; VanderVelde, D.; Powell, D. R.; Bowman-James, K. *J. Am. Chem. Soc.* **2004**, *126*, 12272–12273.

53. Werner, F.; Schneider, H. -J. *Helv. Chim.Acta* **2000***, 83*, 465–478.

54. Isiklan, M.; Pramanik, A.; Fronczek, F. R.; Hossain, M. A. *Acta Cryst.* **2010**, *E66*, o2739–o2740.

55. Hofmeister, F. *Arch. Exp. Pathol. Pharmacol.* **1888***, 24*, 247–260.

# Spectroscopic, Structural, and Theoretical Studies of Halide Complexes with a Urea-based Tripodal Receptor

A urea-based tripodal receptor has been studied for halides by $^1$H NMR spectroscopy, DFT calculations and X-ray crystallography, displaying strong affinity for fluoride anion. The interaction of fluoride anion with the receptor was further confirmed by 2D NOESY and $^{19}$F NMR spectroscopy in DMSO-$d_6$. Crystallographic studies of the chloride, bromide, and silicon hexafluoride complexes of protonated receptor reveals that the anion is externally located via multiple H-bonds.

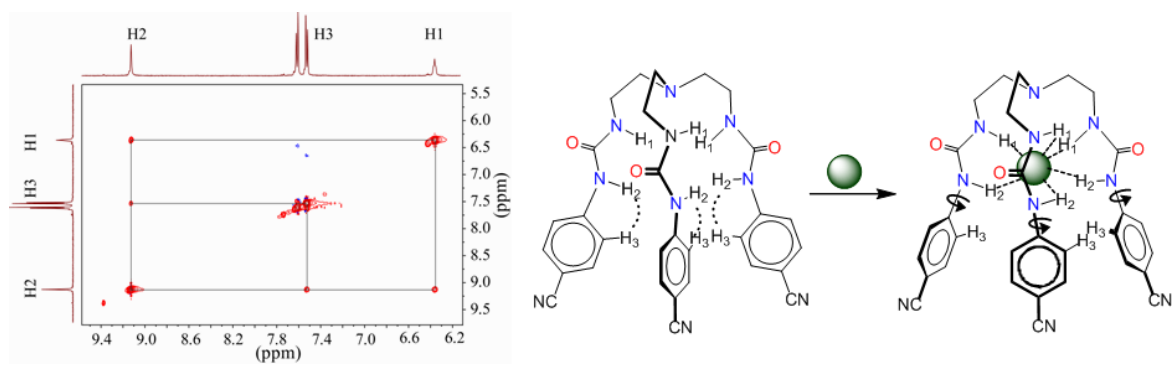